\documentclass[10pt,aps,twocolumn,showpacs,superscriptaddress,groupedaddress,floatfix,nofootinbib,prd]{revtex4-2}
\usepackage{array}
\usepackage{multirow}
\usepackage{graphicx}  
\usepackage{dcolumn}   
\usepackage{bm}        
\usepackage{amssymb}
\usepackage{amsmath}   
\usepackage{comment}
\usepackage{hyperref}  
\usepackage[justification=raggedright, singlelinecheck=false]{caption}

\usepackage{subcaption}
\usepackage{placeins}

\hypersetup{colorlinks,linkcolor={blue},citecolor={blue},urlcolor={red}}
\labelformat{section}{Section #1} 
\labelformat{subsection}{Section #1} 
\labelformat{equation}{Eq.~(#1)} 
\labelformat{figure}{Fig.~#1} 
\labelformat{subfigure}{Fig.~\thefigure#1} 
\labelformat{table}{Tab.~#1} 

\begin{document}
\title{Stability analysis of power-law cosmological models}
\author{Jose Mathew}
\email{E-mail: josemathew@thecochincollege.edu.in}
\affiliation{Department of Physics, The Cochin College, Kochi 682 002, Kerala, India}
\author{A Thariq}
\affiliation{Department of Physics, The Cochin College, Kochi 682 002, Kerala, India}
\begin{abstract} 
In this paper, we revisit the stability of power-law models, focusing on an alternative approach that differs significantly from the standard approaches used in studying power-law models. In the standard approach, stability is studied by reducing the system of background FRW equations to a one-dimensional system for a new background variable $X$ in terms of the number of e-foldings. However, we rewrote the equations, incorporating $H$ into the system and went on to do the calculations up to the second order.
We demonstrate by computing the deviations from the power-law exact solution to second-order in time and show that power-law contraction is never an attractor in time, regardless of parameter values.  Our analysis shows that while first-order corrections align with existing interpretations, second-order corrections introduce significant deviations that cannot be explained by a simple time shift that explains the first-order diverging terms. With importance, we note that in the number of e-folds, the system remains an attractor, while in cosmic time, it is unstable. We also support our claim with numerical results. This new insight has broader implications for the study of attractor behaviour of differential equation solutions and raises questions about the stability of scenarios like the ekpyrotic bounce driven by an exponential potential. Our work also hints that the different temporal variables we use might not be equivalent.
\end{abstract}
\maketitle
\section{Introduction}
Cosmic inflation and the bounce paradigm represent two pivotal concepts in modern cosmology, offering distinct perspectives on the early universe's evolution. Proposed by Alan Guth in 1981~\cite{Guth:1980zm}, cosmic inflation postulates a rapid expansion phase that addresses the horizon problem and the flatness problem. On the other hand, the bounce paradigm challenges the classical Big Bang singularity, proposing a scenario where the universe transitions from a contracting phase to an expanding one without encountering a singularity~\cite{Steinhardt:2001st,Khoury_2001,brandenberger2009matter,bamba2015bounce,LEHNERS2008223}. This concept has profound implications for cosmological models, providing alternative explanations for the universe's origins. Bounce cosmology, though less explored compared to inflation, has garnered significant interest among researchers seeking to understand the early universe's dynamics. Current studies in cosmology emphasize the refinement of both inflationary and bounce models, along with the search for observational evidence to support or refute these paradigms~\cite{Ade2014, Planck2018}. For a comprehensive understanding of cosmological theories and observations, refer to relevant literature and books~\cite{book:15483,book:15486,book:15485,book:75690,martin2014encyclopaedia,baumann2022cosmology,dodelson2020modern,brandenberger2017bouncing,battefeld2015critical}.

One of the early models of inflation is the power-law inflation proposed by  F. Lucchi and  S. Matarrese~\cite{Lucchin:1984yf}. The model assumes the evolution of the scale factor to be power-law ($a\propto t^p$, where $p$ is a positive constant greater than $1$). This is one of the simplest models of inflation that withstood even the latest observational tests~\cite{martin2014encyclopaedia,martin2014best}. Power-law inflation and its spin-offs~\cite{liddle1989power,yokoyama1988dynamics,unnikrishnan2013resurrecting,ohashi2013anisotropic,sharma2022power} are widely studied even today and form the basis of many cosmological models of early Universe, which include models of inflation~\cite{Chiba:2024iia,unnikrishnan2013resurrecting} and cosmic bounce~\cite{Raveendran:2023dst,Nandi:2022twa}, and of late time acceleration~\cite{Kadam:2022yrj}. One of the most essential requirements of early Universe models is their ability to solve fine-tuning problems. Hence, a successful model must have the evolution of the Universe to be an attractor.  For this reason, the stability analysis of both background evolution and the evolution of cosmological perturbations is important.

In this paper, we reexamine the stability of the background solution of the power-law model. Indeed, it is a topic that has been studied widely in depth and breadth~\cite{HALLIWELL1987341,liddle1989power,Finelli_2002,Kallosh_2001,Heard_2002,Arapo_lu_2019,Steinhardt:2001st,Khoury_2001,Ijjas:2024oqn}. However, the earlier studies, where they converted the FRW equations in time to equations in number of e-foldings ($N$) for the two quantities defined as $X={\kappa\dot{\phi}}/{(\sqrt{6}H)}$ and $Y= \kappa\sqrt {V}/(\sqrt{3}H)$~\cite{Heard_2002} missed the point that, in doing so, we might not get the evolution of relevant quantities such as scalar field, Hubble parameter etc., in time.  The evolution of these quantities in the number of e-foldings is not enough to talk about the stability of their evolution. Instead, one must look into the evolution of Hubble parameter, scalar field, \textit{et cetera} in time. Because the number of e-foldings is a function of $H$, the behaviour of $H$ will affect the evolution of the number of e-foldings.

One can note that it is impossible to write these background quantities as a function of the terms $X$ and $Y$. Our analysis, where we consider equations in time, redefines the conditions the parameters must satisfy for a stable expansion and contraction. Previously, there were studies in the literature in this regard, that is, studying the perturbations in time~\cite{Steinhardt:2001st,Khoury_2001}. However, they did the analysis to first-order and came to the conclusion that the perturbation would only result in a shift in time. From our numerical results, we conclude that this is not the case. Further, we analyzed the perturbations to the second order and confirmed our conclusion.

For expansion, that is for an inflationary evolution; though the analysis and results differ, the conclusion regarding the stability of the power-law expansion being an attractor remains the same without any change. However, the conditions for a stable contraction differ drastically. That is, our work is significant for the ekpyrotic bouncing model driven by the exponential potential. Also, our observation that different temporal variables are not completely identical is important.

The outline of this paper is as follows: in the next section, we review the power-law model of inflation and the conditions for its stability. Then, we analytically study the background dynamics, giving importance to the stability of background evolution, where we also compare our numerical results with analytical results. Subsequently,  we show that $X$ and other background quantities are attractors in $N$.  Finally, in conclusion, we review the results and discuss their importance in the current studies of inflation, bouncing models and stability analysis in general.

We define the reduced Planck mass by $M_p=1/\sqrt{8\pi G}$, where $G$ is Newton's gravitational constant and $\kappa=1/M_p$. The sign of the metric is taken to be $(-,+,+,+)$. Latin indices $a,b,c,\dots$ shall be used to run over spacetime indices.

\section{Power-law models}
We shall confine our discussion to cosmological models described by the flat FLRW line element \cite{dodelson2020modern}:  
\begin{equation}
    ds^2=-dt^2+a^2(t)\left[dx^2 + dy^2 + dz^2\right]~.
\end{equation}
Power-law models are those where the scale factor $a(t)$ has the form of a power-law in terms of the cosmic time $t$\footnote{Power-law models in conformal time are also considered in the literature \cite{Boyle:2004gv,Wands:1998yp}. Except for the $p=1$ case, models that are power-law in cosmic time are also power-law in conformal time, while de Sitter expansion is a power-law only in conformal time.}. As is well-known, the matter-dominated and radiation-dominated epochs in the history of the universe are represented by power-law scale factors \cite{dodelson2020modern}. Power-law models have also been used to model inflationary expansion \cite{Lucchin:1984yf,HALLIWELL1987341,liddle1989power} as well as the contracting phase before a bounce \cite{Wands:1998yp,Brandenberger:2012zb,Khoury:2001zk,LEHNERS2008223}.

The scale factor for power-law models takes the form
\begin{align}
    a(t)=a_0\vert  t \vert^p~,
\label{eq:a_power}
    \end{align}
where the constant $a_0$ is positive, the origin of $t$ has been chosen to coincide with $a(t)=0$\footnote{Since we do not expect general relativity to hold close to $a(t)=0$, we situate our discussion far enough from that point} and the constant $p$ may in general be any real number. Neglecting the uninteresting constant scale factor of $p=0$, consider first the case $p>0$. Then, this scale factor will represent an expansion for $t>0$ and a contraction for $t<0$. For $p<0$, it will be the other way around. In this paper, we shall only consider cases with $p>0$.  

The original and simplest power-law model~\cite{Lucchin:1984yf} considers an action of the form
{\small
\begin{equation}
    S=\int d^4x\sqrt{-g}\left[\frac{1}{2\kappa^2}R-\frac{1}{2}g^{a b}\nabla_a\phi\nabla_b\phi-V_0 e^{\lambda\kappa\left(\phi -\phi_0\right)}\right],
\end{equation}
}
where general relativity is minimally coupled to a scalar field with an exponential potential with the constants $V_0$, $\lambda$ and $\phi_0$ as free parameters. The exponential potential allows a smooth inflationary expansion and the cosmological perturbations to have a scale-invariant spectrum. 
where $t$ is the cosmic time and $a(t)$ is the scale factor. The Friedmann equations are given by
\begin{align}
    H^2 &= \frac{\kappa^2}{3} \left( \frac{1}{2} \dot{\phi}^2 + V(\phi) \right)~, \\
    \dot{H} &= -\frac{\kappa^2}{2}  \dot{\phi}^2~.
\end{align}
and the Klein-Gordon equation for the evolution of the scalar field is
\begin{equation}
    \ddot{\phi} + 3H\dot{\phi} + \frac{dV}{d\phi} = 0~.
\end{equation}

Here, $H = {\dot{a}}/{a}$ is the Hubble parameter, representing the rate of expansion of the Universe. The dots denote derivatives with respect to cosmic time \(t\). The power-law solutions\footnote{Note that these are not general solutions with the complete integration constants. Instead these are particular solutions} that satisfy the field equations are~\cite{Lucchin:1984yf}

\begin{subequations}
\begin{equation}
    a(t)=a_0(C\, +\, q_p\, t)^p; 
\end{equation}
\begin{equation}
\phi(t)=-\frac{\sqrt{2\,p}\, \ln \, \left(\mathit{q_p} t +C \right)}{\kappa}+\phi_0
\end{equation}
\label{eq:solns}
    \end{subequations}
    where $C$ and $p$ are two arbitrary constants,  $q_p=\pm1$.  Introducing the constants $C$ and $q_p$ makes the solution more general~\cite{mathew2023starobinsky, Mathew:2016anx}. With these solutions, the overall constant in the potential is fixed as $$V_0=\frac{q_p^2}{\kappa^2}\, p \,\left(3 p -1\right).$$ With, $p=2/\lambda^2$. We can obtain a contracting solution in two ways: either by assuming $q_p$ to be $-1$ and $C$ to take a large positive value, and also here, $t>0$. Now, this can also be achieved by assuming $t$ to take negative values ($-\infty$ to $0$), here again, $q_p=-1$. In the remaining part of the paper, we take $q_p=-1$, $C=0$, and $t$ take negative values for a contracting solution. For an expanding solution $q_p=1$, $C=0$ and $t>0$ are assumed. 
    
Now, let us define $X=\dot{\phi}/\left(\sqrt{6}HM_p\right)$ and $Y=\sqrt{V}/\left(\sqrt{3}H M_p\right)$. For this definition, the field equations take the form

\begin{align}
    \frac{dX}{dN}&=-3X \,Y^2-\lambda\sqrt{\frac{3}{2}} Y^2, \label{eq:sec2eqs1}\\
    1&=X^2+Y^2,
    \label{eq:sec2eqs2}
\end{align}
\begin{equation}
    \frac{dH}{dN}=-3\,X^2\,H,
    \label{eq:dHbydN}
\end{equation}

where $N$ is the number of e-foldings. The above equations lead to a single equation
\begin{equation}
    \frac{dX}{dN}=-3X\left(1-X^2\right)-\lambda\sqrt{\frac{3}{2}}\left(1-X^2\right)=f(X)
    \label{Eq:DEinX}
\end{equation}
Now, we can see that $X=X_p=-\lambda/\sqrt{6}$ and $Y=Y_p=\pm\sqrt{1-\lambda^2/6}$ is a fixed point, i.e., for this point \ref{eq:sec2eqs1} and \ref{eq:sec2eqs2} are satisfied, and the derivatives of both $X$ and $Y$ are zero. If the evolution exactly falls at these critical points, it remains there indefinitely.  Also, these are the very power-law solutions we obtained earlier, \ref{eq:solns}, where
\begin{align}
    H_p&=\frac{p\,q_p}{\left(\mathit{q_p} t +C \right)};\\ \hfill \phi_p&\equiv \phi(t)=-M_p\,{\sqrt{2\,p}\, \ln \, \left(\mathit{q_p} t +C \right)}+\phi_0; \\
    \dot{\phi_p}&=-M_p\,{\sqrt{2\,p}} \frac{q_p}{\left(\mathit{q_p} t +C \right)};\\ V\left(\phi_p\right)&=V_0\,\left(\mathit{q_p} t +C \right)^{-2},\notag\\
    &\;\;\mbox{where}\; {V_0=M_p^2{q_p^2}\, p \,\left(3 p -1\right)}.
    \label{Eq:connectingXandqnties1}
\end{align}

Hence, we have
\begin{align}
    X_p&=\frac{\dot{\phi}}{{\sqrt{6} \,H_p\,M_p}}=-\frac{\lambda}{\sqrt{6}}; \label{Eq:connectingXandqnties2}\\Y_p&= \sqrt{\frac{V}{3}}\frac{1}{H\,M_p}=\pm\, \sqrt{1-\frac{\lambda^2}{6}}.
    \label{Eq:connectingXandqnties3}
\end{align}

We can clearly see that $Y_p$ is imaginary for $\lambda^2>6$.  This is not unexpected as we need a negative value for $V$ in situations like bounce. Also, the value of $Y_p$ is not important for our purpose; in the relevant equations and definitions, we don't have $Y_p$; instead, what appears is $Y_p^2$.

Taking the Taylor expansion of \ref{Eq:DEinX} to the first order, we have
\begin{equation}
    \frac{d(X_{p}+\Delta X)}{dN} = f(X_p)+f'(X)|_{X=X_p}\Delta X+ O(\Delta X^2) 
\end{equation}
We know $dX_p/dN=f(X_p)=0$ and neglecting higher-order terms the equation becomes
\begin{equation}
    \frac{d\Delta X}{dN}=-\left(\frac{6-\lambda^2}{2}\right)\Delta X
    \label{eq:normeq}
\end{equation}
Solving \ref{eq:normeq}, we have
\begin{equation}
\Delta X=C\,e^{-\left(3-\lambda^2/2\right)N}=C\,a^{-\left(3-\lambda^2/2\right)}
    \label{eq:dXint}
\end{equation}
So, if $\lambda^2<6$, the fixed point solutions are attractor solutions. This argument seems elegant. Also, it was verified numerically for inflationary solutions~\cite{HALLIWELL1987341,liddle1989power}.  For contracting solutions, the ekpyrotic case with $\lambda^2>6$ is considered an attractor~\cite{LEHNERS2008223}.  For contracting solutions,  we can argue that if expanding solutions are repellers, the corresponding contracting solutions would be attractors~\cite{Gratton:2003pe}. That is, $\lambda^2>6$ is the required condition for a power-law contraction to be an attractor. In the next section, we revisit the stability analysis.

\section{Stability of power-law solutions}

In the previous section, we obtained fixed-point power-law solutions and the conditions for these solutions to be attractors. However, the quantities that are studied are $X$ and $Y$ and the temporal variable chosen was $N$. But the relevant quantities we have to study in order to see the nature of the evolution of the Universe in this model are $H$, $\phi$ and $\dot{\phi}$ and most importantly, we must study the evolution in time.  Also, the analysis gives the conditions for the solutions $X$ and $Y$, defined in the previous section, to be stable when evolved in $N$. Note that those calculations couldn't explicitly show the evolution of the Hubble parameter and the scalar field in time.  Hence, we can argue that the analysis is not complete.  That is, we don't think the stability of the evolution of solutions $X$ and $Y$ in $N$ guarantees that the evolution of $H$, $\phi$ and other relevant quantities will be stable in time. Her, we proceed to reproduce the solutions for perturbed quantities obtained by~\cite{Steinhardt:2001st}, analytically and then verify them numerically.


In this section, we analyse the stability of the relevant background quantities. For that, let us obtain the differential equations in time,  keeping the variable $X$ intact. For this, we have the equations $$\frac{dX}{dt}=\frac{dX}{dN}\frac{dN}{dt}=H\frac{dX}{dN}$$  and $$\frac{dH}{dt}=-3\,X^{2} H^{2}.$$ Once we have $H$ and $X$, we have all the relevant quantities.

Here, the system of differential equations is
\begin{subequations}
\begin{equation}
   \frac{dX}{dt}= 3HX^{3} + \frac{H\sqrt{6}\lambda X^{2}}{2} - \frac{H\sqrt{6}\lambda}{2} - 3HX = f_1(X,H)
\end{equation}
\begin{equation}
   \frac{dH}{dt}=-3X^2H^2=f_2(X,H)
\end{equation}
   \label{eq:Numeqn}
\end{subequations}
The relevant solutions, as shown in~\ref{Eq:connectingXandqnties2} and \ref{Eq:connectingXandqnties3} are $X=X_p$ and the Hubble parameter, which is not a constant in time but a function of time, $H=H_p=p/t$ and $X=X_p=-\lambda/\sqrt{6}$. 
Now, Taylor expanding about the exact solution to the first order; we have
\begin{widetext}
    \begin{align}
    \frac{d}{dt}[X_p+\Delta X]&=  f_1(X_p,H_p)+\partial_X{f_1(X,H)|_{\{X=X_p,H=H_p\}}}\Delta{X}
    +\partial_H{f_1(X,H)|_{\{X=X_p,H=H_p\}}}\Delta H\\
    \frac{d}{dt}[H_p+\Delta H]&=
    f_2(X_p,H_p)+\partial_X{f_2(X,H)|_{\{X=X_p,H=H_p\}}}\Delta{X}
    +\partial_H{f_2(X,H)|_{\{X=X_p,H=H_p\}}}\Delta H
\end{align}
\end{widetext}
We know that $dX_p/dt=f_1(Xp,H_p)$ and $dH_p/dt=f_2(X_p,H_p)$
Hence, we get the equation
\begin{equation}
{(\Delta x)}^{.}=
\begin{pmatrix}
\Delta{x_1} \\
\Delta{x_2}
\end{pmatrix}^{.}
\approx
\begin{pmatrix}
J_{11} & J_{12} \\
J_{21} & J_{22}
\end{pmatrix}
\begin{pmatrix}
\Delta x_1 \\
\Delta x_2
\end{pmatrix}
\label{eq:xdotjx}
\end{equation}
where,

$\Delta x_1=\Delta X$, $\Delta x_2=\Delta H$ and\hfill\\ 
$J_{ij}=\partial_{x_j}{f_i(x_1,x_2)}|_{\{x_1=X_p,x_2=H_p\}}$
\\\\
 For our solution, the matrix $J$ is given by
 \begin{equation}
 J=
\left[\begin{array}{cc}
\frac{ (\lambda^{2}-6)}{\lambda^{2} t} & 0 \\
 \frac{4 \sqrt{6}}{\lambda^{3}t^{2}} & -\frac{2}{ \,t} 
\end{array}\right]
\label{Eq:J}
\end{equation}

Solving the \ref{eq:xdotjx} using $J$ given in \ref{Eq:J}, we have our solution:
\begin{eqnarray}
    \Delta{X}&=&{D_1}t (-t)^{\frac{-6}{\lambda^{2}}}\label{eq:dxint2}\\
    \Delta{H}&=&\left[\frac{2\, D_1 \sqrt{6} (-t)^{\frac{-6}{\lambda^{2}}} }{\lambda (\lambda^{2}-3)}+ \frac{D_2}{(-p)} \frac{(-p)}{t^2}\right]
    \label{eq:MAIN}
   \end{eqnarray}
   Also, integrating $H$ with respect to $'t'$, we have $N=N_p+\Delta  N=\int{\left(H_p\,+ \Delta H \right)}dt$,
   \begin{equation}
\frac{ N}{N_p}=1+D_1\,\frac{\lambda^3\sqrt{6}\,t\,(-t)^{\frac{-6}{\lambda^2}}}{(\lambda^4-9\lambda^2+18)\,\mbox{ln}(|t|)}- D_2\frac{\lambda^2}{2\,t\, \mbox{ln}(|t|)\,}
\label{eq:Nrep}
\end{equation}

where $N_p$ is the expression for the number of e-folding considering $H=H_p$. 
And, $D_1$ and $D_2$ are integration constants that must be fixed from initial conditions.

Note that in addition to $a_0$ and the integration constant $C$ ($a_0$ is trivial and $C$ could be estimated through the time translation symmetry of FRW equations,  in \ref{eq:solns}.) we have two integration constants and need two initial values to fix them completely. From the above equations, note that the condition for an attractor solution is $\lambda^2<6$ for an expanding Universe. 
However, for a contracting Universe, the situation is tricky. The first term inside the square bracket in the equation for $\Delta H$, \ref{eq:MAIN}, decays for $\lambda^2>6$ in a contracting Universe, and the second term increases; hence, it may appear to diverge. We would like to draw your attention to the fact that this growing term is proportional to $\dot{H}$; that is, this term could be the first term in the Taylor series expansion of H with a time shift $H(t)\approx H_p(t)=p/t\rightarrow H_p(t-D_2/p)=p/(t-D_2/p)$, this term, $-D_2/p$, owes to the shift in time that appears due to the time translational symmetry of FRW metric. Now, if the analogous term appears in all the physical quantities of the system, like $\phi$, $\dot{\phi}$, $N$, \textit{et cetera}, we can confirm that this term is not physical. To confirm this, one has to calculate the first-order perturbation of $\phi$, and it must contain a term (a growing term for our case), which is $-\dot{\phi}\vert_{H=H_p,X=X_p}D_2/p$. And indeed, this is the case for our scenario. This means the big crunch happens at a different time for different initial conditions. The evolution remains the same other than this time shift. Hence, the growing term is not physical. It was shown for the first time in~\cite{Steinhardt:2001st} and was further extended in~\cite{Khoury_2001}. In these papers, the authors calculated the first-order perturbation for the scalar field, $\phi$, and like in our case for the Hubble parameter, they also obtained that there is a growing term. They identified that this term appears due to the time shift, which may be ignored as FRW metrics have time translational symmetry.  However, we noticed that our numerical calculation did not match this conclusion. From numerical computation, we noticed that $H$ is not exactly equal to $H_p(t-D_2/p)$ or evolves to $H_p(t-D_2/p)$ as required. Instead, it diverges away. Hence, we must resort to second-order calculations to resolve the problem. Let us compute the second-order perturbation of $H$ and $X$.

Let us define the Hubble parameter and X to be $H=H_p+\epsilon \, \Delta H+\epsilon^2 \,\Delta^{(2)} H+O(\Delta^{(3)}H)$ and $X=X_p+\epsilon \, \Delta X+\epsilon^2 \, \Delta^{(2)} X+O(\Delta^{(3)}X)$. Then we have the perturbed equations to second-order to be
\begin{align*}
0&=\dot{H}_p  + 3 X_p^2 H_p^2  \\
& + \left(\Delta \dot{H} + 3 \left(2 X_p^2 H_p \Delta H + 2 X_p \Delta X H_p^2 \right)  \right) \epsilon \\
& + \left(\Delta^{(2)} \dot{H} + 3 \left( X_p^2 \left(2 H_p \Delta^{(2)} H + \Delta H^2\right) \right. \right. \\
& \quad \left. \left. + 4 X_p \Delta X H_p \Delta H + \left(2 X_p \Delta^{(2)} X + \Delta X^2\right) H_p^2 \right) \right) \epsilon^2
\end{align*}
and 

\begin{align*}
0&=\dot{X}_p  + \frac{-6 X_p^3 - \lambda \sqrt{6} \left(X_p^2 - 1\right) + 6 X_p}{2} H_p \\
& + \left( \Delta \dot{X} + \frac{-6 X_p^3 - \lambda \sqrt{6} \left(X_p^2 - 1\right) + 6 X_p}{2} \Delta H\, \right. \\
& \quad \left. + \frac{-18 X_p^2 \Delta X - 2 \lambda \sqrt{6} X_p \Delta X + 6 \Delta X}{2} H_p \right) \epsilon \\
& + \left( \Delta^{(2)} \dot{X} + \frac{-6 X_p^3 - \lambda \sqrt{6} \left(X_p^2 - 1\right) + 6 X_p}{2} \Delta^{(2)} H \right. \\
& \quad + \frac{-18 X_p^2 \Delta X - 2 \lambda \sqrt{6} X_p \Delta X + 6 \Delta X}{2} \Delta H \\
& \quad \left. + \frac{-6 \left(X_p \left(2 X_p \Delta^{(2)} X + \Delta X^2\right) + 2 \Delta X^2 X_p \right)}{2\,\kappa} \right. \\
& \quad \left. \left. + \frac{- \sqrt{6} \lambda \left(2 X_p \Delta^{(2)} X + \Delta X^2\right) + 6 \Delta^{(2)} X}{2} H_p \right) \epsilon^2 \right.
\end{align*}
We can separately solve order by order. Solving the second-order equation for $\Delta^{(2)}H$, and assuming $\lambda=-\sqrt{7}$ we get

\begin{align*}
&\Delta^{(2)}H=\\
& \frac{115 D_1^2}{28 (-t)^{5/7}} + \frac{3 D_1 D_2 \sqrt{42}}{14 (-t)^{13/7}} 
- \frac{C_1 \sqrt{42}}{14 (-t)^{6/7}} + \frac{7 D_2^2}{2 (t)^3} + \frac{C_2}{(t)^2} 
\end{align*}

where the integration constants could be fixed from the initial conditions $\Delta X(t_i)$, $\Delta H(t_i)$, which are the deviations of $H(t_i)$ and $X(t_i)$ from $H_p(t_i)$ and $X_p(t_i)$ respectively. Now, without loss of generality, we can take $\Delta^{(2)}X(t_i)=\Delta^{(2)}H(t_i)=0$; here, $\lambda$ is taken to be $-\sqrt{7}$ since the expression would be cumbersome without assuming a value for $\lambda$. Further, we have assumed $\lambda=-\sqrt{7}$ in the numeric calculations.

 So to second order, we get $H$ to take the form:

\begin{align*}
H^{(2)}=\frac{2}{7\,t} &+  \left( -\frac{\sqrt{7} D_1 \sqrt{6}}{14 (-t)^{6/7}} + \frac{D_2}{t^2} \right) \\
&+  \left( \frac{115 D_1^2}{28 (-t)^{5/7}} + \frac{3 D_1 D_2 \sqrt{42}}{14 (-t)^{13/7}} \right. \\
&\quad \left. - \frac{C_1 \sqrt{42}}{14 (-t)^{6/7}} +\frac{7 D_2^2}{2 \,t^3} + \frac{C_2}{t^2} \right)
\end{align*}

Similarly, we have $X$ to the second order given by

\begin{align*}
X^{(2)}&=\sqrt{\frac{7}{6}} 
-  D_1 (-t)^{1/7} \\
&+  \left( \frac{7 D_1^2 (-t)^{2/7} \sqrt{6} \sqrt{7}}{4} - C_1 (-t)^{1/7} 
- \frac{D_1 D_2}{2 (-t)^{6/7}} \right)
\end{align*}
Since, while solving for the integration constants, we consider the first-order solution and second-order solution separately, $D_1$ and $C_1$ are two different constants that can be determined from initial conditions.

We would like to stress the fact that the $FRW$ equations in the conventional form require two integration constants to define the system completely. The equations we use also require two integration constants. Hence, we can confidently say that no additional constraints are there in the system.
 
 In the second-order terms in the Hubble parameter, note the presence of  $7\,D_2^2/2t^3=(-D_2/p)^2 (p/t^3)$, which is the second term in the Taylor series expansion of $H_p(t-D_2)$. Hence, we can argue that this term is not physical. Now, $\Delta^{(2)}H$ contains terms growing faster than $1/t$ which are $C_2/t^2$ and $ {3 D_1 D_2 \sqrt{42}}/{(14 (-t)^{13/7})}$. Here, $C_2/t^2$ is equivalent to the first-order term in the Taylor expansion of $H$. Hence, we can conclude that if time translation alone explains the growing mode in $H$, then $H$ can be approximated to
 
 \[
            H\approx H_c=H_p(t-(D_2+C_2)/p).
            \label{eqHc}
\]
As shown here,  $H$ is closer to the solution $H_c$ compared to $H_p(t-D_2)$, making us guess whether the solution could be explained by a better time-translated solution. We don't think that can happen because it is highly improbable that the term  $ {3 D_1 D_2 \sqrt{42}}/{(14 (-t)^{13/7})}$ will be eliminated by a higher-order term, unlike this term here in the second order, among first-order terms there wasn't a growing mode other than $D_2/t^2$, $D_2/t^2$ could be explained by time translation, and this is the reason why incorporating $C_2$ appearing in second-order perturbed terms, along with $D_2$ explains the solution better. Hence, this will not happen with subsequent orders. That is, the presence of  $ {3 D_1 D_2 \sqrt{42}}/{(14 (-t)^{13/7})}$ in $H$ makes the problem non-trivial. Similarly, $X^{(2)}$ also contains a growing mode that is physical. In the case of $X$, any term containing time ($t$ ) raised to a negative power can be considered a growing term. $X$ is a better variable to capture the stability of the solution because, in the case of $H$ or, for that matter, any other variable with the particular solution that is a function of time, is ailed by the uncertainty in figuring out the correct time translation. New higher-order perturbed terms may be added to the time translation when exploring higher order terms. 

\begin{figure}[h!]
        \includegraphics[width=\linewidth]{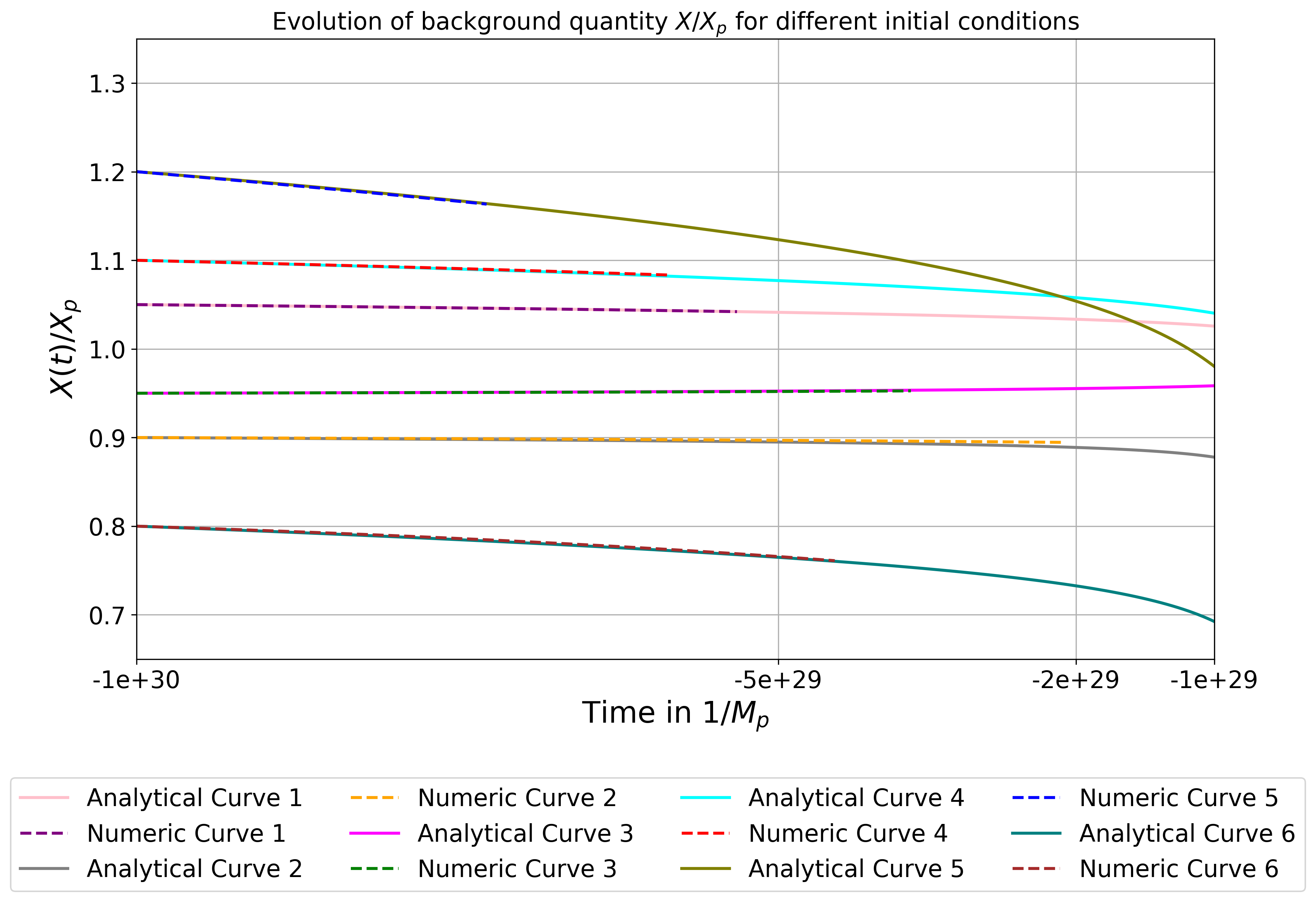}
        \caption{The evolution of $X$ for different initial conditions. We plot $X(t)/X_p$ obtained numerically ($X_{num}/X_p$) and analytically ($X^{(2)}/X_p$) $vs$ $\mbox{Time in}\;1/M_p$. See \ref{tab:curves} for initial conditions and other details. We used the $RKf45$ adaptive step size method available with maple for numerical calculations. Due to the inability to attain sufficient accuracy, the numerical curves break intermittently}
        \label{fig:plotX}
\end{figure}

We also show numerically that the system is unstable when evolving in $t$. See \ref{fig:plotX}, where $X/X_p$ obtained both numerically (We solve equations~ \ref{eq:Numeqn} numerically.) and analytically are plotted against time for different initial conditions, see \ref{tab:curves} for the values of the initial conditions we choose and the values of respective integration constants. The plot clearly demonstrates that the system is unstable when it evolves in time. If it was an attractor, the curves would have evolved towards $1$. But, here, we can see that the curves are repelled away from $1$. In the next section, we show that $X$ is an attractor in $N$. We discuss the implications of these confusing and exciting results in the conclusion. Our results clearly say that the contracting solutions are not attractors.

For numerical results, we had to resort to extremely high accuracy (in maple, setting $\mbox{absolute}\, \mbox{error}=10^{(-22)}$,  $\mbox{relative}\,\mbox{error}=10^{-28}$ and $\mbox{Digits}$, the precision in number of decimal digits, to be $300$). The numerical calculation breaks as the step size effectively reaches zero\footnote{We tried to solve the differential equations numerically using $RK4$, $RK6$ and $RKf45$ adaptive step size method in $C++$, Here, we used arbitrary precision libraries such as $\mbox{BOOST multi-precision lybrary }$, $\mbox{mpfr lybrary}$. \textit{et cetra}. However, we figured out that nothing can match the advanced algorithms of maple, both in speed and precision. Also, whether to use $C++$ or maple depends on the specific problem. It was found that for the evolution in $N$, $C++$ offers a better choice which also helps in understanding the problem better}.  Here, we would like to point out that since the numeric computation breaks by reaching step size effectively zero even with such high precision, we must not resort to numeric computation. It can only be used to guide us in our study. For this reason, we think it's the analytic result that correctly depicts the scenario. Also, note that the integral of $H$ with respect to $t$ gives $N$ and is also a repeller in time. We hope this explains our confusing result.
On the other hand, the inflationary solution with $\lambda^2<6$ is found to be stable. The second-order perturbative analysis also shows no deviation from the standard picture.

So, a power-law contraction can never be an attractor in this setup. This poses serious problems for the ekpyrotic scenario.

\begin{widetext}
\begin{center}
\begin{table}[h!]
\centering
\begin{tabular}{|c|c|c|c|c|c|c|c|}
\hline
Curve & $\Delta H(t_i)$ & $\Delta X(t_i)$ & $D_1$ & $D_2$ & $C_1$ & $C_2$ \\
\hline
1 & $-2.86\times10^{-34}$ & $5.40\times10^{-2}$ &$-2.80\times 10^{-6} $ &$-2.53\times10^{28}$  & $1.68\times 10^{-6}$ & $3.36\times10^{27}$ \\
\hline
2 &$2.86\times10^{-32}$  &$-1.08 \times10^{-1}$&$5.59\times10^{-6}$  & $7.86\times10^{28}$ &$ 6.63\times10^{-6} $& $2.12\times 10^{28}$   \\
\hline
3 &$-2.86\times10^{-32}$  & $-5.40\times10^{-2}$ & $2.80\times10^{-6}$ &$-3.57\times10^{27}$  & $1.72\times 10^{-6}$ &$3.69\times10^{27}$   \\
\hline
4 & $2.86\times10^{-32}$ &  $1.08\times10^{-1}$&  $-5.59\times 10^{-6}$& $-2.14\times10^{28}$ & $6.79\times10^{-6}$ &$1.12\times10^{28}$   \\
\hline
5 & $-2.86\times10^{-35}$ & $2.16\times10^{-1}$ & $-1.11\times10^{-5}$ & $-1.00\times10^{29}$ & $2.69\times10^{-5}$ & $5.33\times10^{28}$   \\
\hline
6 & $2.86\times 10^{-36}$ & $-2.16\times10^{-1}$ & $1.12\times10^{-5}$ & $1.00\times10^{29}$ & $2.69\times 10^{-5}$ &  $5.33\times10^{28}$ \\
\hline
\end{tabular}
\caption{Table showing the initial conditions $\Delta H(t_i)=H(t_i)-H_p(t_i)$ and $\Delta X(t_i)=X(t_i)-X_p(t_i)$ chosen at random but maintaining $H(t_i)$ and $X(t_i)$ near to $H_p(t_i)$ and $X_p$, and the corresponding values of integration constants for each curve in~\ref{fig:plotX}.}
\label{tab:curves}
\end{table}
\end{center}
\end{widetext}
\subsection{Stability of $X$ and $H$ in number of e-foldings}

By rewriting the differential equation~\ref{Eq:DEinX} and solving, we can obtain a relation between $N$ and $X$. This is a complete solution, and the integration constant, $C_3$, could be absorbed into $N$. If any complex value arises from the logarithmic terms, then $C_3$ can be split up into two parts, one to cancel any complex value that arises from logarithmic terms and the other to incorporate the initial deviation in $X$, which could be absorbed into $N$. This, together with the fact that $N$ approaches $-\infty$ as $X$ approaches $X_p$, shows that $X$ is an attractor in $N$.

\begin{widetext}
\begin{align}
N + C_3 = 
-\frac{ \ln \left(X + 1\right)}{ \sqrt{6} \, \lambda - 6}
+ \frac{ \ln \left(X - 1\right)}{ \sqrt{6} \, \lambda + 6}
+ \frac{12 \ln \left(\sqrt{6} \, \lambda + 6 X \right)}
{\left(\sqrt{6} \, \lambda + 6\right) \left(\sqrt{6} \, \lambda - 6\right)}
\end{align}
\end{widetext}

Now, we know that the slow-roll parameter $\epsilon_1$, given by $-\dot{H}/H^2$ can be written in terms of $X$ alone.  That is, $\epsilon_1=3X^2$. Now, as $X$ approaches $X_p$, $\epsilon_1$ approaches $\lambda^2/2$. It's not surprising that this happens when $N\rightarrow -\infty$. But at this point, we have $\epsilon_1=\infty/\infty$ as both $\dot{H}$ and $H$ goes to $\infty$.

Let us also do an order-by-order analysis in $N$ for $X$ and $H$.  Here we perturb the equations~\ref{eq:dHbydN}~and~\ref{Eq:DEinX}.
\begin{widetext}
We get the differential equations for $X$ to be 
\begin{align}
   0=&\left( \frac{d}{dN} \Delta X - \frac{\lambda^2 \Delta X}{2} + 3 \Delta X \right) \epsilon 
+ \left( \sqrt{6} \, \lambda \, \Delta X^2 + \frac{d}{dN} \Delta^{(2)} X - \frac{\lambda^2 \Delta^{(2)} X}{2} + 3 \Delta^{(2)} X \right) \epsilon^2\notag\\
+&\left( 2 \sqrt{6} \, \lambda \, \Delta X \, \Delta^{(2)} X + \frac{d}{dN} \Delta^{(3)} X - 3 \Delta X^3 - \frac{\lambda^2 \Delta^{(3)} X}{2} + 3 \Delta^{(3)} X \right) \epsilon^3\hfill
\end{align}
and the equations for $H$ to be
{\small
\begin{align}
&0=\left( \frac{d}{dN} \Delta H + \frac{\lambda^2 \Delta H}{2} - \frac{2 \sqrt{6} \, \Delta X \, e^{-\frac{N \lambda^2}{2}}}{\lambda} \right) \epsilon \notag \\
&+ \left( \frac{d}{dN} \Delta^{(2)} H + \frac{6 \Delta X^2 \, e^{-\frac{N \lambda^2}{2}}}{\lambda^2} - \sqrt{6} \, \lambda \, \Delta X \, \Delta H + \frac{\lambda^2 \Delta^{(2)} H}{2} - \frac{2 \sqrt{6} \, \Delta^{(2)} X \, e^{-\frac{N \lambda^2}{2}}}{\lambda} \right) \epsilon^2 \notag \\
&+ \left( \frac{d}{dN} \Delta^{(3)} H - \sqrt{6} \, \lambda \, \Delta X \, \Delta^{(2)} H - \sqrt{6} \, \lambda \, \Delta^{(2)} X \, \Delta H + \frac{\lambda^2 \Delta^{(3)} H}{2} + 3 \Delta X^2 \, \Delta H - \frac{2 \sqrt{6} \, \Delta^{(3)} X \, e^{-\frac{N \lambda^2}{2}}}{\lambda} + \frac{12 \Delta X \, \Delta^{(2)} X \, e^{-\frac{N \lambda^2}{2}}}{\lambda^2} \right) \epsilon^3
\end{align}}

Now we have the solution for $X$ till the third order given by

\begin{align}
\Delta X &= c_{1} e^{\frac{(\lambda^2 - 6) N}{2}}, \\
\Delta^{(2)}X &= -\frac{2 c_{1}^{2} \sqrt{6} \, e^{(\lambda^2 - 6) N} \lambda}{\lambda^2 - 6} 
+ e^{\frac{(\lambda^2 - 6) N}{2}} c_{2},  \\
\Delta^{(3)}X &= e^{\frac{(\lambda^2 - 6) N}{2}} c_{3} \quad - \quad\frac{c_{1} \left(e^{\frac{(\lambda^2 - 6) N}{2}}\right)^{2}}{\left(\lambda^2 - 6\right)^{2}} 
\;\; \bigg(4 c_{2} \lambda^{3} \sqrt{6} 
- 27 \, e^{\frac{(\lambda^2 - 6) N}{2}} c_{1}^{2} \lambda^{2} 
 - 24 \lambda c_{2} \sqrt{6} 
+ 18 \, e^{\frac{(\lambda^2 - 6) N}{2}} c_{1}^{2} \bigg),
\label{OBOX}
\end{align}
and $H$, till third order given by

\begin{align}
    \Delta H &= 
\left(\frac{4 \sqrt{6} \, c_{1} e^{\frac{(\lambda^2 - 6) N}{2}}}{(\lambda^2 - 6) \lambda} + c_{4} \right) e^{-\frac{N \lambda^2}{2}} \label{OBOH1},\\
\Delta^{(2)}H&=e^{-\frac{N \lambda^{2}}{2}} c_{5} - \frac{2 \, e^{\frac{(\lambda^2 - 6) N}{2}} e^{-\frac{N \lambda^{2}}{2}}}
{\lambda^{2} (\lambda^{2} - 6)} \quad  \left(-c_{4} \lambda^{3} c_{1} \sqrt{6} 
+ 3 \, e^{\frac{(\lambda^2 - 6) N}{2}} c_{1}^{2} 
- 2 \lambda c_{2} \sqrt{6}\right),
\end{align}

\begin{align}
\Delta^{(3)}H=\frac{{2 \,{\mathrm e}^{-\frac{N \,\lambda^{2}}{2}}}
}{2 \left(\lambda^{2}-6\right)^{2}{\lambda^{2}}}  \left(2 \left(\left(c_{1} c_{5} +c_{2}c_4\right) \lambda^{2}+2 c_{3} \right) \lambda  \left(\lambda^{2}-6\right) \sqrt{6}\, {\mathrm e}^{\frac{\left(\lambda^{2}-6\right) N}{2}}+8 \sqrt{6}\, c_{1}^{3} {\mathrm e}^{\frac{3 \left(\lambda^{2}-6\right) N}{2}} \lambda \right.\nonumber \\\left.-3 \left(\lambda^{2}-6\right) \left(c_{1}  \left(c_{1}c_4 \lambda^{2}+4c_2\right) {\mathrm e}^{\left(\lambda^{2}-6\right) N}-\frac{\lambda^{2} c_{6} \left(\lambda^{2}-6\right)}{3}\right)\right).
\end{align}
\end{widetext}
The above calculations clearly show that $H$ and $X$, and hence the other background quantities are attractors in $N$. Note that the terms that appear to grow in the expression for $H$ faster than $H_p(N)={2 \exp\left(-{N \lambda^{2}}/{2}\right)}/{\lambda^{2}}
$ can be explained by a translation in $N$. These terms are similar to the time translation terms we discussed earlier. Now, in time, the background quantities are not attractors. This is confusing, and we think this happens because, in time, the Universe doesn't follow a contracting path indefinitely, and $N$ doesn't take larger and larger negative values following the power law trajectory. That is, $-N$ might not be a monotonically increasing function in time. Instead, there might be an alternate path in which the Universe evolve. We can see from the expression for $H$ that $N$ is a repeller in time. However, we must admit that we haven't fully understood what is happening here.

\section{Conclusion}

One of the primary reasons the Big Bang cosmology is incomplete is that the model suffers from cosmological puzzles, among which the fine-tuning problems are most important. For this very reason, we want the early Universe models, whether Inflation or bounce, to be attractors. Power-law inflation is one of the most important inflation models because of its simplicity and ability to explain observations, as well as because this model seeds many different cosmological models. There are plenty of cosmological models in the literature motivated by power-law inflation~\cite{LEHNERS2008223,unnikrishnan2013resurrecting,Raveendran:2018yyh,ohashi2013anisotropic,unnikrishnan2014consistency}. Hence, the stability analysis of this model is of utmost importance. And it is a topic that has been extensively studied. The study mainly focuses on converting the differential equations in time to differential equations in $N$, defining new quantities $X=\kappa\dot{\phi}/(\sqrt{6}H)$ and $Y=\kappa\sqrt{V}/(\sqrt{3}H)$, and obtaining the conditions for the stability of these quantities. In this process, one must also check that the stability of $X$ and $Y$ and other background quantities in $N$ ensures the stability of the relevant background quantities in time. 

In this paper, we study the cosmological evolution in time for the relevant background quantities up to the second order; this was explored previously to the first order; see \cite{Steinhardt:2001st,Khoury_2001} for details. We found out that the conditions for a stable evolution turn out to be as expected for inflation. However, they drastically differ for a contracting Universe, which means the bouncing models' stability must be reanalyzed. In the second order, we explicitly showed that there are physical growing modes in $X$ and $H$, which means the system is unstable in time. We conclude that power-law contraction is always a repeller, irrespective of the value of the parameter $\lambda$. Also, we note that the argument of time shift making the growing mode nonphysical~\cite{Steinhardt:2001st} at first order won't save power-law contraction from being unstable; at higher orders, there are growing modes that are physical. Hence, ekpyrotic bounces and other models built based on power-law models must be taken cautiously. Interestingly, we got the background quantities to be attractors in the number of e-foldings and repellers in time. This is confusing, and we must admit that we don't fully understand what is happening here. Because if a solution, when looked at from one variable, is an attractor, it must be an attractor when looked at from any other variable that is monotonically increasing with respect to the first variable. However, it must be noted that in the case of contracting evolution, the number of e-foldings diverges away from the expected evolution and is a repeller in time. This could be one of the possible reasons why we are getting the background quantities to be repellers in time when, at the same time, they are attractors in $N$. We think $N$ is not evolving to sufficiently large negative values that the background quantities are attractors. Also, it could be possible that, in time, the solution is a repeller for a short period, and it will evolve back to the power-law solution. However, looking at the form of the second-order terms, it is highly unlikely.  This result, if true, has significant consequences. Because it hints that all temporal variables cannot be considered equivalent. Most importantly, our paper shows that the way in which stability analysis is done has to be re-examined. We want to add to our statement that our analysis doesn't point out that the previous calculations were wrong; instead, it strengthens the previous calculations and results. As pointed out in earlier works, we got background quantities are attractors in the number of e-foldings during contraction for $\lambda^2>6$. But, by pointing out that the evolution of $H$ and $\phi$ in time are the ones we must look into, and they diverge, our conclusions differ from previous conclusions.

\section*{Acknowledgements and Author Contributions}

JM thanks Krishnamohan Parattu for his valuable suggestions at all stages of the work. 
JM identified the problem from his discussions with AT. He conceived the work and did the calculations; further, AT verified these calculations. KP brought Steinhardt's paper to the notice. JM further proceeded with his calculations based on the paper. Finally, JM wrote the paper. Both the authors verified the calculations and agreed on the results. 

\bibliography{references}

\begin{thebibliography}{43}%
\makeatletter
\providecommand \@ifxundefined [1]{%
 \@ifx{#1\undefined}
}%
\providecommand \@ifnum [1]{%
 \ifnum #1\expandafter \@firstoftwo
 \else \expandafter \@secondoftwo
 \fi
}%
\providecommand \@ifx [1]{%
 \ifx #1\expandafter \@firstoftwo
 \else \expandafter \@secondoftwo
 \fi
}%
\providecommand \natexlab [1]{#1}%
\providecommand \enquote  [1]{``#1''}%
\providecommand \bibnamefont  [1]{#1}%
\providecommand \bibfnamefont [1]{#1}%
\providecommand \citenamefont [1]{#1}%
\providecommand \href@noop [0]{\@secondoftwo}%
\providecommand \href [0]{\begingroup \@sanitize@url \@href}%
\providecommand \@href[1]{\@@startlink{#1}\@@href}%
\providecommand \@@href[1]{\endgroup#1\@@endlink}%
\providecommand \@sanitize@url [0]{\catcode `\\12\catcode `\$12\catcode `\&12\catcode `\#12\catcode `\^12\catcode `\_12\catcode `\%12\relax}%
\providecommand \@@startlink[1]{}%
\providecommand \@@endlink[0]{}%
\providecommand \url  [0]{\begingroup\@sanitize@url \@url }%
\providecommand \@url [1]{\endgroup\@href {#1}{\urlprefix }}%
\providecommand \urlprefix  [0]{URL }%
\providecommand \Eprint [0]{\href }%
\providecommand \doibase [0]{https://doi.org/}%
\providecommand \selectlanguage [0]{\@gobble}%
\providecommand \bibinfo  [0]{\@secondoftwo}%
\providecommand \bibfield  [0]{\@secondoftwo}%
\providecommand \translation [1]{[#1]}%
\providecommand \BibitemOpen [0]{}%
\providecommand \bibitemStop [0]{}%
\providecommand \bibitemNoStop [0]{.\EOS\space}%
\providecommand \EOS [0]{\spacefactor3000\relax}%
\providecommand \BibitemShut  [1]{\csname bibitem#1\endcsname}%
\let\auto@bib@innerbib\@empty
\bibitem [{\citenamefont {Guth}(1981)}]{Guth:1980zm}%
  \BibitemOpen
  \bibfield  {author} {\bibinfo {author} {\bibfnamefont {A.~H.}\ \bibnamefont {Guth}},\ }\href {https://doi.org/10.1103/PhysRevD.23.347} {\bibfield  {journal} {\bibinfo  {journal} {Phys. Rev. D}\ }\textbf {\bibinfo {volume} {23}},\ \bibinfo {pages} {347} (\bibinfo {year} {1981})}\BibitemShut {NoStop}%
\bibitem [{\citenamefont {Steinhardt}\ and\ \citenamefont {Turok}(2002)}]{Steinhardt:2001st}%
  \BibitemOpen
  \bibfield  {author} {\bibinfo {author} {\bibfnamefont {P.~J.}\ \bibnamefont {Steinhardt}}\ and\ \bibinfo {author} {\bibfnamefont {N.}~\bibnamefont {Turok}},\ }\href {https://doi.org/10.1103/PhysRevD.65.126003} {\bibfield  {journal} {\bibinfo  {journal} {Phys. Rev. D}\ }\textbf {\bibinfo {volume} {65}},\ \bibinfo {pages} {126003} (\bibinfo {year} {2002})},\ \Eprint {https://arxiv.org/abs/hep-th/0111098} {arXiv:hep-th/0111098} \BibitemShut {NoStop}%
\bibitem [{\citenamefont {Khoury}\ \emph {et~al.}(2001)\citenamefont {Khoury}, \citenamefont {Ovrut}, \citenamefont {Steinhardt},\ and\ \citenamefont {Turok}}]{Khoury_2001}%
  \BibitemOpen
  \bibfield  {author} {\bibinfo {author} {\bibfnamefont {J.}~\bibnamefont {Khoury}}, \bibinfo {author} {\bibfnamefont {B.~A.}\ \bibnamefont {Ovrut}}, \bibinfo {author} {\bibfnamefont {P.~J.}\ \bibnamefont {Steinhardt}},\ and\ \bibinfo {author} {\bibfnamefont {N.}~\bibnamefont {Turok}},\ }\href {https://doi.org/10.1103/PhysRevD.64.123522} {\bibfield  {journal} {\bibinfo  {journal} {Phys. Rev. D}\ }\textbf {\bibinfo {volume} {64}},\ \bibinfo {pages} {123522} (\bibinfo {year} {2001})},\ \Eprint {https://arxiv.org/abs/hep-th/0103239} {arXiv:hep-th/0103239} \BibitemShut {NoStop}%
\bibitem [{\citenamefont {Brandenberger}(2009)}]{brandenberger2009matter}%
  \BibitemOpen
  \bibfield  {author} {\bibinfo {author} {\bibfnamefont {R.}~\bibnamefont {Brandenberger}},\ }\href {https://doi.org/10.1103/PhysRevD.80.043516} {\bibfield  {journal} {\bibinfo  {journal} {Phys. Rev. D}\ }\textbf {\bibinfo {volume} {80}},\ \bibinfo {pages} {043516} (\bibinfo {year} {2009})},\ \Eprint {https://arxiv.org/abs/0904.2835} {arXiv:0904.2835 [hep-th]} \BibitemShut {NoStop}%
\bibitem [{\citenamefont {Bamba}\ \emph {et~al.}(2015)\citenamefont {Bamba}, \citenamefont {Makarenko}, \citenamefont {Myagky},\ and\ \citenamefont {Odintsov}}]{bamba2015bounce}%
  \BibitemOpen
  \bibfield  {author} {\bibinfo {author} {\bibfnamefont {K.}~\bibnamefont {Bamba}}, \bibinfo {author} {\bibfnamefont {A.~N.}\ \bibnamefont {Makarenko}}, \bibinfo {author} {\bibfnamefont {A.~N.}\ \bibnamefont {Myagky}},\ and\ \bibinfo {author} {\bibfnamefont {S.~D.}\ \bibnamefont {Odintsov}},\ }\href {https://doi.org/10.1088/1475-7516/2015/04/001} {\bibfield  {journal} {\bibinfo  {journal} {JCAP}\ }\textbf {\bibinfo {volume} {04}},\ \bibinfo {pages} {001}},\ \Eprint {https://arxiv.org/abs/1411.3852} {arXiv:1411.3852 [hep-th]} \BibitemShut {NoStop}%
\bibitem [{\citenamefont {Lehners}(2008)}]{LEHNERS2008223}%
  \BibitemOpen
  \bibfield  {author} {\bibinfo {author} {\bibfnamefont {J.-L.}\ \bibnamefont {Lehners}},\ }\href {https://doi.org/10.1016/j.physrep.2008.06.001} {\bibfield  {journal} {\bibinfo  {journal} {Phys. Rept.}\ }\textbf {\bibinfo {volume} {465}},\ \bibinfo {pages} {223} (\bibinfo {year} {2008})},\ \Eprint {https://arxiv.org/abs/0806.1245} {arXiv:0806.1245 [astro-ph]} \BibitemShut {NoStop}%
\bibitem [{\citenamefont {Ade}\ \emph {et~al.}(2014)\citenamefont {Ade} \emph {et~al.}}]{Ade2014}%
  \BibitemOpen
  \bibfield  {author} {\bibinfo {author} {\bibfnamefont {P.~A.~R.}\ \bibnamefont {Ade}} \emph {et~al.} (\bibinfo {collaboration} {BICEP2}),\ }\href {https://doi.org/10.1103/PhysRevLett.112.241101} {\bibfield  {journal} {\bibinfo  {journal} {Phys. Rev. Lett.}\ }\textbf {\bibinfo {volume} {112}},\ \bibinfo {pages} {241101} (\bibinfo {year} {2014})},\ \Eprint {https://arxiv.org/abs/1403.3985} {arXiv:1403.3985 [astro-ph.CO]} \BibitemShut {NoStop}%
\bibitem [{\citenamefont {Aghanim}\ \emph {et~al.}(2020)\citenamefont {Aghanim} \emph {et~al.}}]{Planck2018}%
  \BibitemOpen
  \bibfield  {author} {\bibinfo {author} {\bibfnamefont {N.}~\bibnamefont {Aghanim}} \emph {et~al.} (\bibinfo {collaboration} {Planck}),\ }\href {https://doi.org/10.1051/0004-6361/201833910} {\bibfield  {journal} {\bibinfo  {journal} {Astron. Astrophys.}\ }\textbf {\bibinfo {volume} {641}},\ \bibinfo {pages} {A6} (\bibinfo {year} {2020})},\ \bibinfo {note} {[Erratum: Astron.Astrophys. 652, C4 (2021)]},\ \Eprint {https://arxiv.org/abs/1807.06209} {arXiv:1807.06209 [astro-ph.CO]} \BibitemShut {NoStop}%
\bibitem [{\citenamefont {Liddle}\ and\ \citenamefont {Lyth}(2000)}]{book:15483}%
  \BibitemOpen
  \bibfield  {author} {\bibinfo {author} {\bibfnamefont {A.~R.}\ \bibnamefont {Liddle}}\ and\ \bibinfo {author} {\bibfnamefont {D.~H.}\ \bibnamefont {Lyth}},\ }\href {https://doi.org/10.1017/CBO9781139175180} {\emph {\bibinfo {title} {{Cosmological inflation and large scale structure}}}}\ (\bibinfo  {publisher} {Cambridge University Press},\ \bibinfo {year} {2000})\BibitemShut {NoStop}%
\bibitem [{\citenamefont {Mukhanov}(2005)}]{book:15486}%
  \BibitemOpen
  \bibfield  {author} {\bibinfo {author} {\bibfnamefont {V.}~\bibnamefont {Mukhanov}},\ }\href@noop {} {\emph {\bibinfo {title} {Physical foundations of cosmology}}}\ (\bibinfo  {publisher} {Cambridge University Press},\ \bibinfo {year} {2005})\BibitemShut {NoStop}%
\bibitem [{\citenamefont {Linde}(1990)}]{book:15485}%
  \BibitemOpen
  \bibfield  {author} {\bibinfo {author} {\bibfnamefont {A.~D.}\ \bibnamefont {Linde}},\ }\href@noop {} {\emph {\bibinfo {title} {Particle Physics and Inflationary Cosmology}}},\ Contemporary Concepts in Physics\ (\bibinfo  {publisher} {CRC Press},\ \bibinfo {year} {1990})\BibitemShut {NoStop}%
\bibitem [{\citenamefont {Weinberg}(2008)}]{book:75690}%
  \BibitemOpen
  \bibfield  {author} {\bibinfo {author} {\bibfnamefont {S.}~\bibnamefont {Weinberg}},\ }\href@noop {} {\emph {\bibinfo {title} {Cosmology}}}\ (\bibinfo  {publisher} {Oxford University Press, USA},\ \bibinfo {year} {2008})\BibitemShut {NoStop}%
\bibitem [{\citenamefont {Martin}\ \emph {et~al.}(2014{\natexlab{a}})\citenamefont {Martin}, \citenamefont {Ringeval},\ and\ \citenamefont {Vennin}}]{martin2014encyclopaedia}%
  \BibitemOpen
  \bibfield  {author} {\bibinfo {author} {\bibfnamefont {J.}~\bibnamefont {Martin}}, \bibinfo {author} {\bibfnamefont {C.}~\bibnamefont {Ringeval}},\ and\ \bibinfo {author} {\bibfnamefont {V.}~\bibnamefont {Vennin}},\ }\href {https://doi.org/10.1016/j.dark.2014.01.003} {\bibfield  {journal} {\bibinfo  {journal} {Phys. Dark Univ.}\ }\textbf {\bibinfo {volume} {5-6}},\ \bibinfo {pages} {75} (\bibinfo {year} {2014}{\natexlab{a}})},\ \Eprint {https://arxiv.org/abs/1303.3787} {arXiv:1303.3787 [astro-ph.CO]} \BibitemShut {NoStop}%
\bibitem [{\citenamefont {Baumann}(2022)}]{baumann2022cosmology}%
  \BibitemOpen
  \bibfield  {author} {\bibinfo {author} {\bibfnamefont {D.}~\bibnamefont {Baumann}},\ }\href@noop {} {\emph {\bibinfo {title} {Cosmology}}}\ (\bibinfo  {publisher} {Cambridge University Press},\ \bibinfo {year} {2022})\BibitemShut {NoStop}%
\bibitem [{\citenamefont {Dodelson}\ and\ \citenamefont {Schmidt}(2020)}]{dodelson2020modern}%
  \BibitemOpen
  \bibfield  {author} {\bibinfo {author} {\bibfnamefont {S.}~\bibnamefont {Dodelson}}\ and\ \bibinfo {author} {\bibfnamefont {F.}~\bibnamefont {Schmidt}},\ }\href@noop {} {\emph {\bibinfo {title} {Modern cosmology}}}\ (\bibinfo  {publisher} {Academic press},\ \bibinfo {year} {2020})\BibitemShut {NoStop}%
\bibitem [{\citenamefont {Brandenberger}\ and\ \citenamefont {Peter}(2017)}]{brandenberger2017bouncing}%
  \BibitemOpen
  \bibfield  {author} {\bibinfo {author} {\bibfnamefont {R.}~\bibnamefont {Brandenberger}}\ and\ \bibinfo {author} {\bibfnamefont {P.}~\bibnamefont {Peter}},\ }\href {https://doi.org/10.1007/s10701-016-0057-0} {\bibfield  {journal} {\bibinfo  {journal} {Found. Phys.}\ }\textbf {\bibinfo {volume} {47}},\ \bibinfo {pages} {797} (\bibinfo {year} {2017})},\ \Eprint {https://arxiv.org/abs/1603.05834} {arXiv:1603.05834 [hep-th]} \BibitemShut {NoStop}%
\bibitem [{\citenamefont {Battefeld}\ and\ \citenamefont {Peter}(2015)}]{battefeld2015critical}%
  \BibitemOpen
  \bibfield  {author} {\bibinfo {author} {\bibfnamefont {D.}~\bibnamefont {Battefeld}}\ and\ \bibinfo {author} {\bibfnamefont {P.}~\bibnamefont {Peter}},\ }\href {https://doi.org/10.1016/j.physrep.2014.12.004} {\bibfield  {journal} {\bibinfo  {journal} {Phys. Rept.}\ }\textbf {\bibinfo {volume} {571}},\ \bibinfo {pages} {1} (\bibinfo {year} {2015})},\ \Eprint {https://arxiv.org/abs/1406.2790} {arXiv:1406.2790 [astro-ph.CO]} \BibitemShut {NoStop}%
\bibitem [{\citenamefont {Lucchin}\ and\ \citenamefont {Matarrese}(1985)}]{Lucchin:1984yf}%
  \BibitemOpen
  \bibfield  {author} {\bibinfo {author} {\bibfnamefont {F.}~\bibnamefont {Lucchin}}\ and\ \bibinfo {author} {\bibfnamefont {S.}~\bibnamefont {Matarrese}},\ }\href {https://doi.org/10.1103/PhysRevD.32.1316} {\bibfield  {journal} {\bibinfo  {journal} {Phys. Rev. D}\ }\textbf {\bibinfo {volume} {32}},\ \bibinfo {pages} {1316} (\bibinfo {year} {1985})}\BibitemShut {NoStop}%
\bibitem [{\citenamefont {Martin}\ \emph {et~al.}(2014{\natexlab{b}})\citenamefont {Martin}, \citenamefont {Ringeval}, \citenamefont {Trotta},\ and\ \citenamefont {Vennin}}]{martin2014best}%
  \BibitemOpen
  \bibfield  {author} {\bibinfo {author} {\bibfnamefont {J.}~\bibnamefont {Martin}}, \bibinfo {author} {\bibfnamefont {C.}~\bibnamefont {Ringeval}}, \bibinfo {author} {\bibfnamefont {R.}~\bibnamefont {Trotta}},\ and\ \bibinfo {author} {\bibfnamefont {V.}~\bibnamefont {Vennin}},\ }\href {https://doi.org/10.1088/1475-7516/2014/03/039} {\bibfield  {journal} {\bibinfo  {journal} {JCAP}\ }\textbf {\bibinfo {volume} {03}},\ \bibinfo {pages} {039}},\ \Eprint {https://arxiv.org/abs/1312.3529} {arXiv:1312.3529 [astro-ph.CO]} \BibitemShut {NoStop}%
\bibitem [{\citenamefont {Liddle}(1989)}]{liddle1989power}%
  \BibitemOpen
  \bibfield  {author} {\bibinfo {author} {\bibfnamefont {A.~R.}\ \bibnamefont {Liddle}},\ }\href {https://doi.org/10.1016/0370-2693(89)90776-4} {\bibfield  {journal} {\bibinfo  {journal} {Phys. Lett. B}\ }\textbf {\bibinfo {volume} {220}},\ \bibinfo {pages} {502} (\bibinfo {year} {1989})}\BibitemShut {NoStop}%
\bibitem [{\citenamefont {Yokoyama}\ and\ \citenamefont {Maeda}(1988)}]{yokoyama1988dynamics}%
  \BibitemOpen
  \bibfield  {author} {\bibinfo {author} {\bibfnamefont {J.}~\bibnamefont {Yokoyama}}\ and\ \bibinfo {author} {\bibfnamefont {K.-i.}\ \bibnamefont {Maeda}},\ }\href {https://doi.org/10.1016/0370-2693(88)90880-5} {\bibfield  {journal} {\bibinfo  {journal} {Phys. Lett. B}\ }\textbf {\bibinfo {volume} {207}},\ \bibinfo {pages} {31} (\bibinfo {year} {1988})}\BibitemShut {NoStop}%
\bibitem [{\citenamefont {Unnikrishnan}\ and\ \citenamefont {Sahni}(2013)}]{unnikrishnan2013resurrecting}%
  \BibitemOpen
  \bibfield  {author} {\bibinfo {author} {\bibfnamefont {S.}~\bibnamefont {Unnikrishnan}}\ and\ \bibinfo {author} {\bibfnamefont {V.}~\bibnamefont {Sahni}},\ }\href {https://doi.org/10.1088/1475-7516/2013/10/063} {\bibfield  {journal} {\bibinfo  {journal} {JCAP}\ }\textbf {\bibinfo {volume} {10}},\ \bibinfo {pages} {063}},\ \Eprint {https://arxiv.org/abs/1305.5260} {arXiv:1305.5260 [astro-ph.CO]} \BibitemShut {NoStop}%
\bibitem [{\citenamefont {Ohashi}\ \emph {et~al.}(2013)\citenamefont {Ohashi}, \citenamefont {Soda},\ and\ \citenamefont {Tsujikawa}}]{ohashi2013anisotropic}%
  \BibitemOpen
  \bibfield  {author} {\bibinfo {author} {\bibfnamefont {J.}~\bibnamefont {Ohashi}}, \bibinfo {author} {\bibfnamefont {J.}~\bibnamefont {Soda}},\ and\ \bibinfo {author} {\bibfnamefont {S.}~\bibnamefont {Tsujikawa}},\ }\href {https://doi.org/10.1103/PhysRevD.88.103517} {\bibfield  {journal} {\bibinfo  {journal} {Phys. Rev. D}\ }\textbf {\bibinfo {volume} {88}},\ \bibinfo {pages} {103517} (\bibinfo {year} {2013})},\ \Eprint {https://arxiv.org/abs/1310.3053} {arXiv:1310.3053 [hep-th]} \BibitemShut {NoStop}%
\bibitem [{\citenamefont {Sharma}\ and\ \citenamefont {Verma}(2022)}]{sharma2022power}%
  \BibitemOpen
  \bibfield  {author} {\bibinfo {author} {\bibfnamefont {A.~K.}\ \bibnamefont {Sharma}}\ and\ \bibinfo {author} {\bibfnamefont {M.~M.}\ \bibnamefont {Verma}},\ }\href {https://doi.org/10.3847/1538-4357/ac3ed7} {\bibfield  {journal} {\bibinfo  {journal} {Astrophys. J.}\ }\textbf {\bibinfo {volume} {926}},\ \bibinfo {pages} {29} (\bibinfo {year} {2022})}\BibitemShut {NoStop}%
\bibitem [{\citenamefont {Chiba}\ and\ \citenamefont {Houri}(2024)}]{Chiba:2024iia}%
  \BibitemOpen
  \bibfield  {author} {\bibinfo {author} {\bibfnamefont {T.}~\bibnamefont {Chiba}}\ and\ \bibinfo {author} {\bibfnamefont {T.}~\bibnamefont {Houri}},\ }\href {https://doi.org/10.1088/1361-6382/ad7185} {\bibfield  {journal} {\bibinfo  {journal} {Class. Quant. Grav.}\ }\textbf {\bibinfo {volume} {41}},\ \bibinfo {pages} {19LT01} (\bibinfo {year} {2024})},\ \Eprint {https://arxiv.org/abs/2404.19162} {arXiv:2404.19162 [gr-qc]} \BibitemShut {NoStop}%
\bibitem [{\citenamefont {Raveendran}\ and\ \citenamefont {Chakraborty}(2024)}]{Raveendran:2023dst}%
  \BibitemOpen
  \bibfield  {author} {\bibinfo {author} {\bibfnamefont {R.~N.}\ \bibnamefont {Raveendran}}\ and\ \bibinfo {author} {\bibfnamefont {S.}~\bibnamefont {Chakraborty}},\ }\href {https://doi.org/10.1007/s10714-024-03242-8} {\bibfield  {journal} {\bibinfo  {journal} {Gen. Rel. Grav.}\ }\textbf {\bibinfo {volume} {56}},\ \bibinfo {pages} {55} (\bibinfo {year} {2024})},\ \Eprint {https://arxiv.org/abs/2302.02584} {arXiv:2302.02584 [astro-ph.CO]} \BibitemShut {NoStop}%
\bibitem [{\citenamefont {Nandi}\ and\ \citenamefont {Kaur}(2022)}]{Nandi:2022twa}%
  \BibitemOpen
  \bibfield  {author} {\bibinfo {author} {\bibfnamefont {D.}~\bibnamefont {Nandi}}\ and\ \bibinfo {author} {\bibfnamefont {M.}~\bibnamefont {Kaur}},\ }\href@noop {} {\  (\bibinfo {year} {2022})},\ \Eprint {https://arxiv.org/abs/2206.08335} {arXiv:2206.08335 [astro-ph.CO]} \BibitemShut {NoStop}%
\bibitem [{\citenamefont {Kadam}\ \emph {et~al.}(2023)\citenamefont {Kadam}, \citenamefont {Levi~Said},\ and\ \citenamefont {Mishra}}]{Kadam:2022yrj}%
  \BibitemOpen
  \bibfield  {author} {\bibinfo {author} {\bibfnamefont {S.~A.}\ \bibnamefont {Kadam}}, \bibinfo {author} {\bibfnamefont {J.}~\bibnamefont {Levi~Said}},\ and\ \bibinfo {author} {\bibfnamefont {B.}~\bibnamefont {Mishra}},\ }\href {https://doi.org/10.1142/S0219887823500834} {\bibfield  {journal} {\bibinfo  {journal} {Int. J. Geom. Meth. Mod. Phys.}\ }\textbf {\bibinfo {volume} {20}},\ \bibinfo {pages} {2350083} (\bibinfo {year} {2023})},\ \Eprint {https://arxiv.org/abs/2210.17075} {arXiv:2210.17075 [gr-qc]} \BibitemShut {NoStop}%
\bibitem [{\citenamefont {Halliwell}(1987)}]{HALLIWELL1987341}%
  \BibitemOpen
  \bibfield  {author} {\bibinfo {author} {\bibfnamefont {J.}~\bibnamefont {Halliwell}},\ }\href {https://doi.org/https://doi.org/10.1016/0370-2693(87)91011-2} {\bibfield  {journal} {\bibinfo  {journal} {Physics Letters B}\ }\textbf {\bibinfo {volume} {185}},\ \bibinfo {pages} {341} (\bibinfo {year} {1987})}\BibitemShut {NoStop}%
\bibitem [{\citenamefont {Finelli}\ and\ \citenamefont {Brandenberger}(2002)}]{Finelli_2002}%
  \BibitemOpen
  \bibfield  {author} {\bibinfo {author} {\bibfnamefont {F.}~\bibnamefont {Finelli}}\ and\ \bibinfo {author} {\bibfnamefont {R.}~\bibnamefont {Brandenberger}},\ }\bibfield  {journal} {\bibinfo  {journal} {Physical Review D}\ }\textbf {\bibinfo {volume} {65}},\ \href {https://doi.org/10.1103/physrevd.65.103522} {10.1103/physrevd.65.103522} (\bibinfo {year} {2002})\BibitemShut {NoStop}%
\bibitem [{\citenamefont {Kallosh}\ \emph {et~al.}(2001)\citenamefont {Kallosh}, \citenamefont {Kofman},\ and\ \citenamefont {Linde}}]{Kallosh_2001}%
  \BibitemOpen
  \bibfield  {author} {\bibinfo {author} {\bibfnamefont {R.}~\bibnamefont {Kallosh}}, \bibinfo {author} {\bibfnamefont {L.}~\bibnamefont {Kofman}},\ and\ \bibinfo {author} {\bibfnamefont {A.}~\bibnamefont {Linde}},\ }\bibfield  {journal} {\bibinfo  {journal} {Physical Review D}\ }\textbf {\bibinfo {volume} {64}},\ \href {https://doi.org/10.1103/physrevd.64.123523} {10.1103/physrevd.64.123523} (\bibinfo {year} {2001})\BibitemShut {NoStop}%
\bibitem [{\citenamefont {Heard}\ and\ \citenamefont {Wands}(2002)}]{Heard_2002}%
  \BibitemOpen
  \bibfield  {author} {\bibinfo {author} {\bibfnamefont {I.~P.~C.}\ \bibnamefont {Heard}}\ and\ \bibinfo {author} {\bibfnamefont {D.}~\bibnamefont {Wands}},\ }\href {https://doi.org/10.1088/0264-9381/19/21/309} {\bibfield  {journal} {\bibinfo  {journal} {Classical and Quantum Gravity}\ }\textbf {\bibinfo {volume} {19}},\ \bibinfo {pages} {5435–5447} (\bibinfo {year} {2002})}\BibitemShut {NoStop}%
\bibitem [{\citenamefont {Arapoğlu}\ and\ \citenamefont {Yükselci}(2019)}]{Arapo_lu_2019}%
  \BibitemOpen
  \bibfield  {author} {\bibinfo {author} {\bibfnamefont {A.~S.}\ \bibnamefont {Arapoğlu}}\ and\ \bibinfo {author} {\bibfnamefont {A.~E.}\ \bibnamefont {Yükselci}},\ }\href {https://doi.org/10.1142/s021773231950069x} {\bibfield  {journal} {\bibinfo  {journal} {Modern Physics Letters A}\ }\textbf {\bibinfo {volume} {34}},\ \bibinfo {pages} {1950069} (\bibinfo {year} {2019})}\BibitemShut {NoStop}%
\bibitem [{\citenamefont {Ijjas}\ \emph {et~al.}(2024)\citenamefont {Ijjas}, \citenamefont {Steinhardt}, \citenamefont {Garfinkle},\ and\ \citenamefont {Cook}}]{Ijjas:2024oqn}%
  \BibitemOpen
  \bibfield  {author} {\bibinfo {author} {\bibfnamefont {A.}~\bibnamefont {Ijjas}}, \bibinfo {author} {\bibfnamefont {P.~J.}\ \bibnamefont {Steinhardt}}, \bibinfo {author} {\bibfnamefont {D.}~\bibnamefont {Garfinkle}},\ and\ \bibinfo {author} {\bibfnamefont {W.~G.}\ \bibnamefont {Cook}},\ }\href {https://doi.org/10.1088/1475-7516/2024/07/077} {\bibfield  {journal} {\bibinfo  {journal} {JCAP}\ }\textbf {\bibinfo {volume} {07}},\ \bibinfo {pages} {077}},\ \Eprint {https://arxiv.org/abs/2404.00867} {arXiv:2404.00867 [gr-qc]} \BibitemShut {NoStop}%
\bibitem [{\citenamefont {Boyle}\ \emph {et~al.}(2004)\citenamefont {Boyle}, \citenamefont {Steinhardt},\ and\ \citenamefont {Turok}}]{Boyle:2004gv}%
  \BibitemOpen
  \bibfield  {author} {\bibinfo {author} {\bibfnamefont {L.~A.}\ \bibnamefont {Boyle}}, \bibinfo {author} {\bibfnamefont {P.~J.}\ \bibnamefont {Steinhardt}},\ and\ \bibinfo {author} {\bibfnamefont {N.}~\bibnamefont {Turok}},\ }\href {https://doi.org/10.1103/PhysRevD.70.023504} {\bibfield  {journal} {\bibinfo  {journal} {Phys. Rev. D}\ }\textbf {\bibinfo {volume} {70}},\ \bibinfo {pages} {023504} (\bibinfo {year} {2004})},\ \Eprint {https://arxiv.org/abs/hep-th/0403026} {arXiv:hep-th/0403026} \BibitemShut {NoStop}%
\bibitem [{\citenamefont {Wands}(1999)}]{Wands:1998yp}%
  \BibitemOpen
  \bibfield  {author} {\bibinfo {author} {\bibfnamefont {D.}~\bibnamefont {Wands}},\ }\href {https://doi.org/10.1103/PhysRevD.60.023507} {\bibfield  {journal} {\bibinfo  {journal} {Phys. Rev. D}\ }\textbf {\bibinfo {volume} {60}},\ \bibinfo {pages} {023507} (\bibinfo {year} {1999})},\ \Eprint {https://arxiv.org/abs/gr-qc/9809062} {arXiv:gr-qc/9809062} \BibitemShut {NoStop}%
\bibitem [{\citenamefont {Brandenberger}(2012)}]{Brandenberger:2012zb}%
  \BibitemOpen
  \bibfield  {author} {\bibinfo {author} {\bibfnamefont {R.~H.}\ \bibnamefont {Brandenberger}},\ }\href@noop {} {\  (\bibinfo {year} {2012})},\ \Eprint {https://arxiv.org/abs/1206.4196} {arXiv:1206.4196 [astro-ph.CO]} \BibitemShut {NoStop}%
\bibitem [{\citenamefont {Khoury}\ \emph {et~al.}(2002)\citenamefont {Khoury}, \citenamefont {Ovrut}, \citenamefont {Steinhardt},\ and\ \citenamefont {Turok}}]{Khoury:2001zk}%
  \BibitemOpen
  \bibfield  {author} {\bibinfo {author} {\bibfnamefont {J.}~\bibnamefont {Khoury}}, \bibinfo {author} {\bibfnamefont {B.~A.}\ \bibnamefont {Ovrut}}, \bibinfo {author} {\bibfnamefont {P.~J.}\ \bibnamefont {Steinhardt}},\ and\ \bibinfo {author} {\bibfnamefont {N.}~\bibnamefont {Turok}},\ }\href {https://doi.org/10.1103/PhysRevD.66.046005} {\bibfield  {journal} {\bibinfo  {journal} {Phys. Rev. D}\ }\textbf {\bibinfo {volume} {66}},\ \bibinfo {pages} {046005} (\bibinfo {year} {2002})},\ \Eprint {https://arxiv.org/abs/hep-th/0109050} {arXiv:hep-th/0109050} \BibitemShut {NoStop}%
\bibitem [{\citenamefont {Mathew}(2023)}]{mathew2023starobinsky}%
  \BibitemOpen
  \bibfield  {author} {\bibinfo {author} {\bibfnamefont {J.}~\bibnamefont {Mathew}},\ }\href@noop {} {\  (\bibinfo {year} {2023})},\ \Eprint {https://arxiv.org/abs/2307.01899} {arXiv:2307.01899 [gr-qc]} \BibitemShut {NoStop}%
\bibitem [{\citenamefont {Mathew}\ and\ \citenamefont {Shankaranarayanan}(2016)}]{Mathew:2016anx}%
  \BibitemOpen
  \bibfield  {author} {\bibinfo {author} {\bibfnamefont {J.}~\bibnamefont {Mathew}}\ and\ \bibinfo {author} {\bibfnamefont {S.}~\bibnamefont {Shankaranarayanan}},\ }\href {https://doi.org/10.1016/j.astropartphys.2016.07.004} {\bibfield  {journal} {\bibinfo  {journal} {Astropart. Phys.}\ }\textbf {\bibinfo {volume} {84}},\ \bibinfo {pages} {1} (\bibinfo {year} {2016})},\ \Eprint {https://arxiv.org/abs/1602.00411} {arXiv:1602.00411 [astro-ph.CO]} \BibitemShut {NoStop}%
\bibitem [{\citenamefont {Gratton}\ \emph {et~al.}(2004)\citenamefont {Gratton}, \citenamefont {Khoury}, \citenamefont {Steinhardt},\ and\ \citenamefont {Turok}}]{Gratton:2003pe}%
  \BibitemOpen
  \bibfield  {author} {\bibinfo {author} {\bibfnamefont {S.}~\bibnamefont {Gratton}}, \bibinfo {author} {\bibfnamefont {J.}~\bibnamefont {Khoury}}, \bibinfo {author} {\bibfnamefont {P.~J.}\ \bibnamefont {Steinhardt}},\ and\ \bibinfo {author} {\bibfnamefont {N.}~\bibnamefont {Turok}},\ }\href {https://doi.org/10.1103/PhysRevD.69.103505} {\bibfield  {journal} {\bibinfo  {journal} {Phys. Rev. D}\ }\textbf {\bibinfo {volume} {69}},\ \bibinfo {pages} {103505} (\bibinfo {year} {2004})},\ \Eprint {https://arxiv.org/abs/astro-ph/0301395} {arXiv:astro-ph/0301395} \BibitemShut {NoStop}%
\bibitem [{\citenamefont {Raveendran}\ and\ \citenamefont {Sriramkumar}(2019)}]{Raveendran:2018yyh}%
  \BibitemOpen
  \bibfield  {author} {\bibinfo {author} {\bibfnamefont {R.~N.}\ \bibnamefont {Raveendran}}\ and\ \bibinfo {author} {\bibfnamefont {L.}~\bibnamefont {Sriramkumar}},\ }\href {https://doi.org/10.1103/PhysRevD.99.043527} {\bibfield  {journal} {\bibinfo  {journal} {Phys. Rev. D}\ }\textbf {\bibinfo {volume} {99}},\ \bibinfo {pages} {043527} (\bibinfo {year} {2019})},\ \Eprint {https://arxiv.org/abs/1809.03229} {arXiv:1809.03229 [astro-ph.CO]} \BibitemShut {NoStop}%
\bibitem [{\citenamefont {Unnikrishnan}\ and\ \citenamefont {Shankaranarayanan}(2014)}]{unnikrishnan2014consistency}%
  \BibitemOpen
  \bibfield  {author} {\bibinfo {author} {\bibfnamefont {S.}~\bibnamefont {Unnikrishnan}}\ and\ \bibinfo {author} {\bibfnamefont {S.}~\bibnamefont {Shankaranarayanan}},\ }\href {https://doi.org/10.1088/1475-7516/2014/07/003} {\bibfield  {journal} {\bibinfo  {journal} {JCAP}\ }\textbf {\bibinfo {volume} {07}},\ \bibinfo {pages} {003}},\ \Eprint {https://arxiv.org/abs/1311.0177} {arXiv:1311.0177 [astro-ph.CO]} \BibitemShut {NoStop}%
\end{thebibliography}%
\bibliographystyle{apsrev4-2}
\end{document}